%\hsize=17truecm \vsize=25truecm\hoffset=-1.cm
\input epsf.sty
\hfuzz=0.3cm
\tolerance=400
\raggedbottom
\def\hb{\hfill\break}
\font\abst=cmr8
\rightline{FTUAM 01-04, \quad February, 2001}
\rightline{hep-ph/0102312}
\bigskip
\hrule height 0.3mm
\bigskip
\centerline{\uppercase{ Disagreement between standard model and
experiment for  muon g-2 ?}}
\bigskip
\hrule

\vskip0.5truecm
\setbox9=\vbox{\hsize65mm {\noindent\bf F. J. Yndur\'ain} 
\vskip 0.1truecm
\noindent{\sl Departamento de F\'{\i}sica Te\'orica, C-XI,\hb
 Universidad Aut\'onoma de Madrid,}\hb
 Canto Blanco,\quad
E-28049, Madrid, Spain.\hb
\quad({\abst e-mail: fjy@delta.ft.uam.es})
}
\smallskip
\centerline{\box9}
\bigskip
\noindent{\bf Abstract}. {\abst In a recent experimental paper 
reporting a precise measurement of the muon anomalous magnetic moment, Brown et al. claim
disagreement  of their result with evaluations based on the standard model. 
This claim is based on comparison with a single
theoretical  evaluation. It is pointed out that other equally
legitimate  theoretical evaluations exist in the literature, with which the new experimental figure 
is quite compatible. The claim of indications
 for new physics beyond the standard model does  not appear
justified.
In this version we take into account corrections to one of the works quoted, 
we include a new discussion at the end,  
and also minor errors are corrected. The conclusions remain unchanged.}
\vskip0.5truecm
In a recent paper$^{[1]}$ a new measurement is reported for the 
value of the muon anomalous magnetic moment ($g-2$), 
$a_\mu$. After averaging with older determinations one has then the precise {\sl experimental}
result,
$$10^{11}\times a_\mu(\hbox{experiment})=116\,592\,023\pm140\pm60.\eqno{(1)}$$
Brown et al. then compare this with a theoretical evaluation of the same quantity 
and  conclude that there is discrepancy at the 2.7 standard 
deviations level, 
and thus indication of new physics. 
This has triggered a deluge of non-standard model explanations of the `discrepancy', 
of which here we only mention ref.~2.

In order to clarify the issue, we first very briefly review the theoretical calculation 
of $a_\mu$. According to Hughes and Kinoshita$^{[3]}$ we have 
the following theoretical contributions, apart from that of hadronic 
vacuum polarization (h.v.p.), that we discuss later:  
$$\eqalign{10^{11}\times a_\mu(\hbox{Pure QED})= &116\,584\,705.7\pm1.8\cr
10^{11}\times a_\mu(\hbox{Weak})= & 151\pm4. \cr
10^{11}\times a_\mu(\hbox{Hadronic light by light})= &-79.2\pm15.4\cr
10^{11}\times a_\mu(\hbox{Other higher order hadronic})=&-101\pm6\cr
}$$
Excluding the hadronic contributions to vacuum polarization, and 
composig errors quadratically, we then have 
$$10^{11}\times a_\mu(\hbox{Except h.v.p.})= 116\,584\,677\pm17.\eqno{(2)}$$

There is reasonable consensus on these contributions, although I personally 
think the errors on the hadronic pieces to be somewhat underestimated; for example, 
the authors of ref.~2 give 
$$10^{11}\times a_\mu(\hbox{Hadronic light by light})= -86\pm25,$$
but we will use the figure given by Hughes and Kinoshita here.
  
We now we
pass to the more  controversial hadronic 
vacuum polarization which, moreover, provides the bulk of the 
{\sl theoretical} error. There are a number of evaluations
of the h.v.p. corrections; we quote
now four recent ones:
$$\eqalign{10^{11}\times a_\mu(\hbox{h.v.p.})= &6\,924\pm62\quad \hbox{(DH)}\cr
10^{11}\times a_\mu(\hbox{h.v.p.})= &7\,250\pm158\quad \hbox{(EJ)}\cr
10^{11}\times a_\mu(\hbox{h.v.p.})= &6\,988\pm111\quad \hbox{(J)}\cr
10^{11}\times a_\mu(\hbox{h.v.p.})= &7\,113\pm103\quad \hbox{(AY)}\cr}
\eqno{(3)}$$
The meaning of this is: DH stands for the analysis of ref.~4, 
EJ for that of ref.~5 and J for that, as yet unpublished, of F. Jegerlehner, 
quoted in ref.~2, and private communication. 
Finally, AY indicates the result of ref.~6. 

The figure quoted for EJ contains an error in the sense that 
it contains a string of hadron corrections to the 
photon propagator and thus leads to double counting 
when including higher hadronic corrections. So we will 
here consider only the figure denoted by J.

Adding (2) and (3), the theoretical predictions are then,
$$\eqalign{10^{11}\times a_\mu(\hbox{theory})= &116\,591\,601\pm(\sqrt{17^2+62^2}=67)\quad
\hbox{(DH)}\cr
10^{11}\times a_\mu(\hbox{theory})= &116\,591\,665\pm(\sqrt{17^2+111^2}=112)\quad
\hbox{(J)}\cr
10^{11}\times a_\mu(\hbox{theory})= &116\,591\,790\pm(\sqrt{17^2+103^2}=104)\quad
\hbox{(AY)};\cr
10^{11}\times a_\mu(\hbox{theory})= &116\,591\,904\pm(\sqrt{17^2+112^2}=115)\quad
\hbox{(CLY)};\cr
10^{11}\times a_\mu(\hbox{exp.})= &116\,592\,023\pm(\sqrt{140^2+60^2}=152).\cr}
\eqno{(4)}$$
In the  line before last (CLY) we have included the `old' result of ref.~7,
corrected for the new favoured value of higher
 order hadronic corrections,\footnote{*}{If we had taken the value actually reported in CLY, 
eq.~1.4, we would have had a figure that, perhaps by chance, falls dead on top of the 
new experimental result:
$116\,592\,024\pm(\sqrt{27^2+112^2}=115)\;\hbox{(CLY, uncorrected)}$.}
 and 
in the last line we have repeated the 
recent experimental value of ref.~1, to facilitate the comparison.

Subtracting from (1) we find that the `discrepancies' between theory and experiment are 
thus:
$$\eqalign{10^{11}\times\Delta a_\mu(\hbox{exp -- th})=&422\pm152 (\hbox{exp.})\pm77
(\hbox{th})\quad 
\hbox{(DH)}\cr
10^{11}\times\Delta a_\mu(\hbox{exp -- th})=&358\pm152 (\hbox{exp.})\pm112 (\hbox{th})\quad 
\hbox{(J)}\cr
10^{11}\times\Delta a_\mu(\hbox{exp -- th})=&233\pm152 (\hbox{exp.})\pm104 (\hbox{th})\quad 
\hbox{(AY)}\cr
10^{11}\times\Delta a_\mu(\hbox{exp -- th})=&119\pm152 (\hbox{exp.})\pm115 (\hbox{th})\quad 
\hbox{(CLY)}\cr}
\eqno{(5)}$$
with, I believe, self-explanatory notation (we have included the new values of the 
higher hadronic corrections for the CLY figure). 
The first  theoretical evaluation (DH) is more distant from experiments than what 
the errors, at 1$\sigma$ level, would allow. 
J is slightly outside the $1\sigma$ region, and CLY and AY are perfectly compatible with experiment
 (considering
that the evaluations  CLY, EJ and AY antedate the experiment, one should perhaps say that 
experiment validates the standard model theory). 
We will discuss very briefly DH, CLY and AY as representative calculations: each is based on a
different method of evaluation.

All these evaluations use essentially old (pre-1985) experimental data in the critical
$s^{1/2}\leq 5\; GeV$ region for 
$e^+e^-\to\hbox{hadrons}$ since 1985, with the exception of what one can get 
indirectly from the hadronic decays of the $\tau$. 
The variation of DH with respect to CLY is to use these $\tau$ decay results to 
supplement low energy 
 $e^+e^-\to\hbox{hadrons}$ data. 
The improvement of CLY (and AY) with respect to older calculations lies 
in the use of dispersive methods, that allow use of $\pi\pi$ scattering phase shifts, 
and data on the pion form factor in the spacelike region,  
so curbing systematic errors of $e^+e^-\to\hbox{hadrons}$ (at low energy). 
These systematic errors are very important, in particular at `high' energy 
$s^{1/2}\geq 1.5\; GeV$; in some regions, as large as 10\%, as demonstrated e.g. in ref.~8
comparing  various experiments with one another and
 with  QCD predictions. 
Because of these systematic errors, AY use a QCD calculation, plus resonances, above 
$s^{1/2}=1.5\; GeV$. This is what lowers a little the result of AY as
compared to 
CLY.

To summarize: for the h.v.p. contributions, EJ use $e^+e^-\to\hbox{hadrons}$ 
data essentially; AY supplement this by  $\pi\pi$ 
phase shifts and data for the pion form factor in the spacelike region
 (so does CLY) at low eergy,
and  QCD at high energies (CLY uses experimental data at higher energies). Finally, DH 
take $\tau$ decay data to improve the evaluation for low energy, 
and use essentially  experimental
$e^+e^-\to\hbox{hadrons}$  data at higher energies. 
Although based on different methods, all four determinations are compatible
 with one another, within
errors.  The value reported by DH is slightly {\sl lower} than the other three, 
and indeed than almost any other evaluation (see, for example, fig.~3 
in Davier's review, ref.~9, or fig.~1 here).
 
On the face of the results of eq.~(5), what one is tempted to conclude is that experiment favours
the  evaluations of CLY, AY which are in fact in perfect agreement among themselves and 
in good agreement with the new experimental result. This is seen very clearly in the figure~1 
(where I have added also the 
result of Brown and Worstell (BW), ref.~10, just to show that there are other evaluations that
those by myself and my collaborators that are in strict agreement with the new data).
\topinsert{ 
\setbox0=\vbox{\epsfxsize 9.5truecm\epsfbox{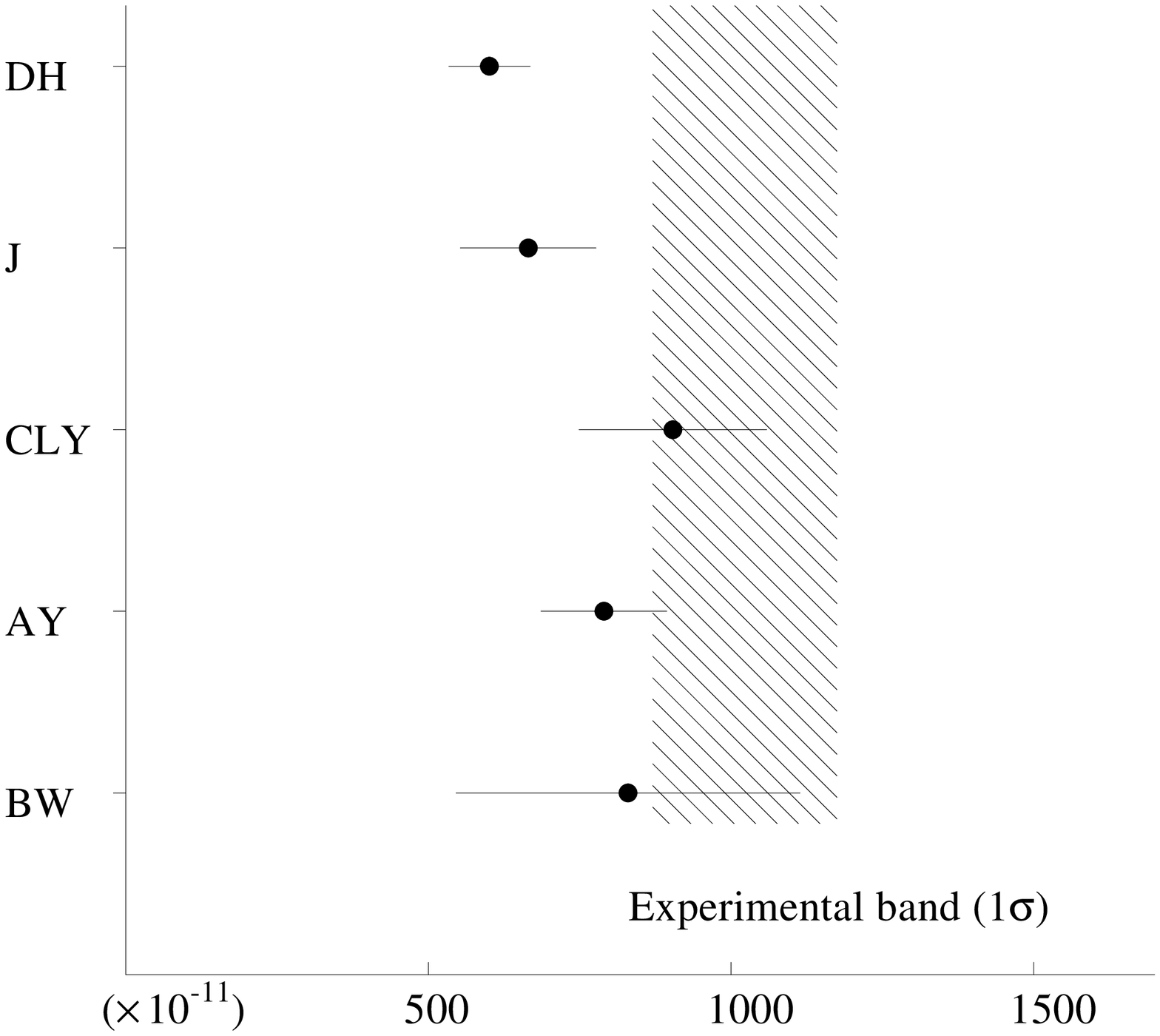}}
\centerline{\box0}
\bigskip
\centerline{{\bf Fig. 1}.- Some theoretical results on the h.v.p., and experiment.} 
\centerline{The quantity plotted is $a_\mu-116\,591\,000\times10^{-11}$.}
\medskip
}\endinsert
Of course, such a conclusion would not be totally correct; 
the discrepancy between the theoretical result of DH and 
experiment is not really large enough to discriminate. 
But, by the same token, it follows that, 
unless one can prove that the evaluation  of Davier and H\"ocker is clearly
superior to the others, and until 
substantially more precise 
theoretical evaluations and experimental data become available, what is completely  unjustified is
to claim any disagreement between  theory and experiment. 

To advertise evidence for SUSY or any other kind of
 {\sl nonstandard physics} on such 
basis is, to put 
things in as mild a way as possible, misleading.
\bigskip
\noindent{\sl Further discussion}. It has been claimed in a number of 
places, in particular by some of the propounders of the ``harbinger of new physics" 
interpretation, that the result of DH is the best because
it gives smallest errors.  Unfortunately, however, small errors are not always a sign of improved
evaluation; many times they just reflect unjustified  optimism. 
The fact that the error in DH is a factor about 2 or more smaller than 
{\sl all} the others, including the very recent evaluation of 
Jegerlehner (J) that uses all 
data available to DH, and a few more, 
should make one suspicious about the claims of 
accuracy made by Davier and  H\"ocker.  
\bigskip

\noindent{\bf Acknowledgements} I am grateful to W.~Lucha and B.~Gavela who 
brought this problem to my attention, 
and to several other members of the Department in UAM and
IFT for  discussions. Correspondence with F.~Jegerlehner, who 
clarified some points of his analysis is acknowledged, as is 
a remark of M.~Davier that allowed me to correct a typo in the 
previous version of this note.

 Thanks are also due
to CICYT for financial support.

\bigskip
\noindent{\bf References}
{
\item{1}{H.N. Brown et al., hep-ex/0102017.}
\item{2}{A. Czarnecki and W. J. Marciano, hep-ph/0102122 v2.}
\item{3}{V. W. Hughes and T. Kinoshita, Rev. Mod. Phys., {\bf 71}, S133 (1999).}
\item{4}{M. Davier and A. H\"ocker, Phys. Letters {\bf B435}, 427 (1998).}
\item{5}{S. Eidelmann and F. Jegerlehner, Z. Phys. {\bf C67}, 585 (1995).}
\item{6}{K. Adel and F. J. Yndur\'ain, Rev. Acad. Ciencias (Esp.), {\bf 92} (1998) 
[hep-ph/9509378]}.
\item{7}{A. Casas, C. L\'opez and F. J. Yndur\'ain, Phys. Rev. {\bf 32}, 736 (1985).}
\item{8}{R. Marshall, Z. Phys. {\bf C43}, 595 (1989).}
\item{9}{M. Davier, Proc. 5th Symposium on 
Tau Lepton Physics, Santander, hep-ph/9812370 v2.}
\item{10}{D. H. Brown and W. A. Worstell, Phys. Rev. {\bf D54}, 3237 (1996).}
\item{}{}
}
\bye